\documentclass[11pt,a4paper]{article}
\usepackage[inactive, tightpage]{preview}
\usepackage{jheppub}
\usepackage{amsthm}
\usepackage[T1]{fontenc}
\usepackage[utf8]{inputenc}
\usepackage{mathtools}   
\usepackage{amsfonts}
\usepackage{physics}
\usepackage{mathrsfs}
\usepackage{natbib}
\usepackage{comment}
\usepackage{ytableau}
\usepackage{tikz, pgfplots}
\usepackage{ytableau}
\usetikzlibrary{patterns, calc, patterns, angles, quotes, decorations, shadows}
\usepackage{enumitem}
\usepackage{subcaption}
\usepackage[font={small}]{caption}
\usepackage{graphicx}
\usepackage{amssymb}
\usepackage{microtype}
\usepackage{svg} 
\usepackage{todonotes}

\usepackage{microtype}
\usepackage{hyperref}

\definecolor{darkmidnightblue}{rgb}{0.0, 0.2, 0.4}
\hypersetup{
    colorlinks = true,
    linkcolor=darkmidnightblue,
    citecolor=darkmidnightblue
}

\usepackage{cleveref}

\usepackage{booktabs}
\usepackage{pdflscape}
\usepackage{afterpage}
\usepackage{flafter}
\usepackage{rotating}
\usepackage{fancyhdr}
\usepackage{slashed}
\fancypagestyle{lscape}{%
\fancyhf{} 
\fancyfoot[LE]{}
\fancyfoot[LO] {}

}

\newcommand{\Z}{\mathbb{Z}}
\newcommand{\C}{\mathbb{C}} 

\DeclareMathSymbol{:}{\mathord}{operators}{"3A}

\theoremstyle{definition}

\title{WZW Partition Functions from Supersymmetric Localization}

\author{Boan Zhao}

\affiliation{DAMTP, University of Cambridge, Cambridge, United Kingdom}

\emailAdd{bz258@cam.ac.uk}

\abstract{
We prove a conjecture of Murthy and Witten \cite{MurthyWitten} which expresses diagonal modular invariant WZW partition functions as lattice sums \footnote{The ideas and the techniques used in this paper already appeared in \cite{Warner1989,  Roberts1990}}.
}
\begin{document}
\maketitle
\section{Introduction}
Conformal field theories (CFTs) in two dimensions (2d) form a large class of exactly solvable quantum field theories. The partition function of a 2d CFT on a complex torus with modularity parameter $\tau$ must be invariant under the modular transformations $\tau \to \tau + 1, \tau \to -1/\tau$ \cite{DiFrancesco, Cardy1986}. In the context of $SU(N)$ Wess-Zumino-Witten (WZW) models at level $k$ \cite{WZW_original_paper}, the simplest modular invariant partition function is the diagonal invariant:
\begin{eqnarray}\label{eq: diagonal_inv}
    Z_{\text{diag}}(q, \bar{q}, x, y) \coloneq \sum_{\lambda\in \text{integrable}} \chi^{\widehat{SU(N)}_k}_\lambda(q, x)\chi^{\widehat{SU(N)}_k}_{\mu}(\bar{q}, y)
\end{eqnarray}
where $q = \exp(2\pi i \tau), \bar{q} = \exp(-2\pi i \bar{\tau})$ and $\chi_{\lambda}^{\widehat{SU(N)}_k}(q, x)$ are integrable characters of the affine Lie algebra $\widehat{SU(N)}_k$ labelled by representation $\lambda$ of $SU(N)$. The vectors $x = (x_1, x_2,..., x_N), y = (y_1, y_2,..., y_N)$ are the fugacities for the $SU(N) \times SU(N)$ symmetries of the model and they satisfy $\prod_{i = 1}^Nx_i = \prod_{i=1}^N y_i = 1$. 

Recently, motivated by results from supersymmetric localization, Murthy and Witten \cite{MurthyWitten} studied the relationship between a particular lattice sum\footnote{Our overall normalization constant may differ from the one in \cite{MurthyWitten}. We choose the normalization constant so that it is cancelled after the Poisson resummation. The subscript susy indicates that it comes from a computation in supersymmetric localisation.}:
\begin{eqnarray}\label{eq:lattice_sum_schem}
    Z_{\text{susy}}(\tau, \bar{\tau}, \alpha_L, \alpha_R) = \sqrt{\frac{2\tau_2}{k + N}}^{1 - N}|\Lambda_R^*|\nonumber\sum_{\substack{w\in W\\n, s\in \Lambda_R}}\det(w)
    \exp(-\frac{\pi(k + N)}{\tau_2}f(n, s, \tau, \bar{\tau}, w(\alpha_L), \alpha_R))
\end{eqnarray}
and the diagonal invariant \eqref{eq: diagonal_inv}. Here $\alpha_L = (\alpha_L^1,..., \alpha_L^N), \alpha_R = (\alpha_R^1,..., \alpha_R^N)$ are both vectors of length $N$. $\tau_2$ is the imaginary part of $\tau$. The function $f$ will be defined in the first part of chapter \ref{ch:diagonal}. $W\cong S_N$ is the Weyl group of $SU(N)$ and $\det(w)$ is the determinant of $w$ as a orthogonal linear map on the Cartan subalgebra of $SU(N)$. $\Lambda_R$ is the root lattice of $SU(N)$ and $|\Lambda_R^*|$ is the volume of the unit cell of the dual lattice. If $w\in W$, $w(\alpha_L)$ is the natural permutation action of the symmetric group $S_N$ on vectors. Murthy and Witten conjectured that the lattice sum equals the numerator of the diagonal modular invariant:
\begin{eqnarray}
    \frac{Z_{\text{susy}}(\tau, \bar{\tau}, \alpha_L, \alpha_R)}{\text{free fermions}} = \sum_{\lambda} \chi_\lambda^{\widehat{SU(N)}_k}(q, x)\chi_\lambda^{\widehat{SU(N)}_k}(\bar{q}, y) 
\end{eqnarray}
under the identification
\begin{eqnarray}
    q = \exp(2\pi i \tau)\quad \bar{q} = \exp(-2\pi i \bar{\tau})\quad x_i = \exp(2\pi i \alpha_R^i)\quad y_i = \exp(-2\pi i \alpha_L^i)
\end{eqnarray}
The denominator is a product of free fermion partition functions transforming in the adjoint of $SU(N)$. An explicit formula for the denominator will be given later. Murthy and Witten proved this conjecture in the case of $N = 2$ and we will generalize their proof for all $N\geq 2$.
\section{Proof of the Murthy-Witten Conjecture}\label{ch:diagonal}
In this chapter, we prove that the Murthy-Witten conjecture \eqref{eq:lattice_sum_schem}. First we set out conventions (see appendix \ref{app: convention} for more details). $\alpha_L, \alpha_R$ are complex vectors whose components sum to zero:
\begin{eqnarray}
    \alpha_L = (\alpha_L^1,..., \alpha_L^N)\quad \sum_i \alpha_L^i = 0\\ \alpha_R = (\alpha_R^1,..., \alpha_R^N)\quad \sum_i \alpha_R^i = 0
\end{eqnarray}
We use $SU(N)$ fugacities $x = (x_1,..., x_N), y = (y_1,..., y_N)$ related to $\alpha_L, \alpha_R$ in the following way:
\begin{eqnarray}
     x_i = \exp(2\pi i \alpha_R^i)\quad y_i = \exp(-2\pi i \alpha_L^i)\Rightarrow \prod_{i = 1}^N x_i = \prod_{i = 1}^N y_i = 1
\end{eqnarray}
We also write
\begin{eqnarray}
    \tau = \tau_1 + i\tau_2\quad q = \exp(2\pi i \tau)\quad \bar{q} = \exp(-2\pi i \tau)
\end{eqnarray}
We assume that $\tau_2 > 0$ so $|q| < 1$.

Characters $\chi_\mu^{\widehat{SU(N)}_k}$ of $\widehat{SU(N)}_k$ have the following representation due to Weyl and Kac \cite{DiFrancesco, Kac1990}:
\begin{eqnarray}
    \chi_\lambda^{\widehat{SU(N)}_k}(q, x) = \frac{\sum_{w\in W}\det(w)\theta_{w(\lambda + \rho)}(q, x)}{Z_{\text{fermi, adj}}(q, x)}
\end{eqnarray}
where $w$ sums over the Weyl group of $SU(N)$ and theta functions are defined in appendix \ref{app: convention}. The Weyl vector is the following element of the complexified Cartan subalgebra of $SU(N)$:
\begin{eqnarray}
    \rho = \left(\frac{N - 1}{2}, \frac{N-3}{2},..., \frac{1- N}{2}\right)
\end{eqnarray}
The denominator is the free fermion partition function in the adjoint of $SU(N)$:
\begin{equation}
    Z_{\text{fermi, adj}}(q, x) = (q;q)_\infty^{-1} q^{\rho^2/(2N)}\prod_{N\geq i > j\geq 1}\left(\sqrt{\frac{x_j}{x_i}} - \sqrt{\frac{x_i}{x_j}}\right) \prod_{i, j = 1}^N \left(q\frac{x_i}{x_j};q\right)_\infty
\end{equation}
where $\rho$ is defined in appendix \ref{app: convention} and $\rho^2$ uses the inner product defined in \ref{app: convention}.
$Z_{\text{fermi, adj}}$ is independent of $\lambda$. The Pochhammer symbol is defined as
\begin{equation}
    (a; q)_\infty = \prod_{n = 0}^\infty (1 - aq^n)\quad a\in \C, |q| < 1
\end{equation}
Murthy and Witten introduced the following lattice sum:
\begin{align}\label{eq:lattice_sum}
    Z_{\text{susy}}(\tau, \bar{\tau}, \alpha_L, \alpha_R) := 
    \sum_{\substack{w\in W\\n, s\in \Lambda_R}}\det(w)
    \exp(-\frac{\pi(k + N)}{2\tau_2}f(n, s, \tau, \bar{\tau}, w(\alpha_L), \alpha_R)) \sqrt{\frac{2\tau_2}{k + N}}^{1 - N}|\Lambda^*|\nonumber
\end{align}
where\footnote{The $-2i\tau_2(n, s)$ term (or equivalently $(-1)^{(k + N)(n, s)}$ after taking out of the exponential) is due to Yongchao Lu. It comes from evaluating the WZW term
\begin{eqnarray}
    S_{\text{WZW}}(g) = \int_{M_3}\frac{1}{12\pi} \epsilon^{\mu\nu \rho}\tr(g^{-1}\partial_\mu g g^{-1}\partial_\nu g \partial_\rho g)d^3x
\end{eqnarray}
on the BPS locus which consists of (up to an overall multiplication by an element of $SU(N)$) a group homomorphism $g: T^2 \to SU(N)$ with image in the maximal torus of $SU(N)$ (diagonal matrices). $M_3$ is a three-manifold whose boundary is $T^2$ and we extend $g$ from $T^2$ to $M_3$ (which is always possible as $SU(N)$ is simply connected). To compute the WZW term, we write $g = g_1g_2$ where $g_1, g_2$ are the restrictions of $g$ to the two independent one-cycles of $T^2$. $g_1, g_2$ are extended to the whole $T^2$ by composing with the natural projection maps to the two cycles. The WZW terms of $g_1, g_2$ are both zero as we can set $M_3 = S^1 \times D^2$ and the $D^2$ bounds the cycle which $g_i$ maps to a point. So the image of $S^1 \times D^2$ is one-dimensional in $SU(N)$ and pullback of the biinvariant three-form is zero. The rest of the computation is done via the formula \cite{DiFrancesco}
\begin{eqnarray}
    S_{\text{WZW}}(g_1g_2) = S_{\text{WZW}}(g_1) + S_{\text{WZW}}(g_2) + \int_{T^2} \frac{1}{4\pi}\epsilon^{\mu\nu}\tr(g_1^{-1}\partial_\mu g_1 g_2^{-1}\partial_\nu g_2)d^2x
\end{eqnarray}
We now assume that the worldsheet $T^2$ is a square $[0,1]\times [0, 1]$ where the opposite edges are identified. We use $x^1, x^2$ as the two real coordinates along the two intervals. Then we set $g_1^{-1}\partial_1 g_1 = 2\pi in, g_2^{-1}\partial_2g_2 = 2\pi is, \partial_2 g_1 = \partial_1 g_2 = 0$, where $n,s$ are elements of the root lattice (vectors of length $N$ which sum to zero). They can be identified with diagonal matrices whose diagonal entries are the two vectors. The trace $\tr(ns)$ becomes the Euclidean inner product $(n, s)$ and we have $S_{\text{WZW}}(g_1, g_2) = -\pi (n, s)$. In the path integral we need to insert $\exp(i(k + N)S_{\text{WZW}}(g_1g_2))$ which becomes the $(-1)^{-(k + N)(n, s)} = (-1)^{(k + N)(n, s)}$ factor, since $(n, s)$ is always an integer.
We thank Noah Porcelli for discussions on this argument.}
\begin{eqnarray}
    f(n, s, \tau, \bar{\tau}, \alpha_L, \alpha_R) =  (n - s\bar{\tau} - 2 \alpha_L , n - s\tau + 2 \alpha_R) -2i\tau_2(n, s) +  (\alpha_L + \alpha_R)^2
\end{eqnarray}
Here $W$ is the Weyl group of $SU(N)$. $\Lambda_R$ is the root lattice of $SU(N)$. The complex bilinear inner product $(\cdot, \cdot)$ between vectors is the usual Euclidean inner product (appendix \ref{app: convention}). \footnote{This definition coincides with the original definition by Murthy and Witten only when $\alpha_{L, R}$ are real. We would like our partition functions to depend holomorphically on $\alpha_L, \alpha_R$ and so we have performed an analytic continuation from the real locus.}. 

Murthy and Witten conjectured that:
\begin{eqnarray}
    \frac{Z_{\text{susy}}(\tau, \bar{\tau}, \alpha_L, \alpha_R)}{Z_{\text{fermi, adj}}(q, x)Z_{\text{fermi, adj}}(\bar{q}, y)} = \sum_{\lambda\in \text{integrable}}\chi_\lambda^{\widehat{SU(N)}_k}(q, x)\chi_\lambda^{\widehat{SU(N)}_k}(\bar{q}, y)
\end{eqnarray}
where $\lambda$ sums over integrable weights of $SU(N)$ at level $k$ (appendix \ref{app: convention}).

We will prove the equivalent statement:
\begin{eqnarray}\label{eq: conjecture}
    \begin{aligned}
    &\sum_{\lambda\in \text{integrable}}\left(\sum_{w\in W}\theta_{w(\lambda + \rho)}(q, x)\det(w)\right)\left(\sum_{w'\in W}\theta_{w'(\lambda + \rho)}(\bar{q}, y)\det(w')\right)\\
    = & \sum_{\substack{w\in W\\ n, s\in \Lambda_R}}\det(w) (-1)^{(k + N)(n, s)} \sqrt{\frac{2\tau_2}{k + N}}^{1 - N}|\Lambda_R^*|\\
    &\quad\quad\times \exp(-\frac{\pi(k + N)}{2\tau_2}((n - s\bar{\tau} - 2 w(\alpha_L) )(n - s\tau + 2 \alpha_R )
    + (w(\alpha_L) + \alpha_R)^2))
    \end{aligned}
\end{eqnarray}
The proof will be divided into two steps. During the first step, we rearrange the sums of products of theta functions in a different way. During the second step, we perform a Poisson resummation of the lattice sum $Z_{\text{susy}}$.
\subsection{Weyl Folding of Theta Functions}
First we convert the sum over $\lambda\in \text{integrable}$ to $\lambda \in \Lambda_W/(k + N)\Lambda_R$, the quotient of the weight lattice of $SU(N)$ by $(k+ N)$ times the root lattice of $SU(N)$.
\begin{align*}
    &\sum_{\lambda\in \text{integrable}}\left(\sum_{w\in W}\theta_{w(\lambda + \rho)}(q, x)\det(w)\right)\left(\sum_{w'\in W}\theta_{w'(\lambda + \rho)}(\bar{q}, y)\det(w')\right)\\
    = &\frac{1}{|W|}\sum_{\lambda\in \Lambda_W/(k + N)\Lambda_R}\left(\sum_{w\in W}\theta_{w(\lambda + \rho)}(q, x)\det(w)\right)\left(\sum_{w'\in W}\theta_{w'(\lambda + \rho)}(\bar{q}, y)\det(w')\right)
\end{align*}
where $|W| = N!$ is the size of the Weyl group of $SU(N)$. To prove this formula, consider the shifted Weyl action on $\Lambda_W/(k + N)\Lambda_R$:
\begin{eqnarray}
    \lambda \mapsto w(\lambda + \rho) - \rho\quad \lambda \in \Lambda_W/(k+N) \Lambda_R\quad w\in W
\end{eqnarray}
It is known \cite{DiFrancesco} that
\begin{enumerate}\label{enum: shifted_weyl_action}
    \item The shifted Weyl orbit of an integrable weight $\lambda$ has size $|W|$. In other words, the stabilizer of an integrable weight under the shifted Weyl action is trivial.
    \item Any other weight $\lambda\in \Lambda_W$ outside these orbits is fixed by a shifted Weyl transformation of determinant -1. In this case the corresponding sum
    \begin{eqnarray*}
        \sum_{w\in W}\theta_{w(\lambda + \rho)}(q, x)\det(w)
    \end{eqnarray*} is zero.
\end{enumerate}
A proof of these facts can be found in appendix \ref{sec:shifted_weyl}. The product
\begin{eqnarray*}
    \left(\sum_{w\in W}\theta_{w(\lambda + \rho)}(q, x)\det(w)\right)\left(\sum_{w'\in W}\theta_{w'(\lambda + \rho)}(\bar{q}, y)\det(w')\right)
\end{eqnarray*}
is constant on each shifted Weyl orbit as both terms pick up the determinant of the shifted Weyl transformation. Therefore, we need to divide by $1/|W|$. The proof of the formula is complete.

We can further simplify the formula as follows:
\begin{equation}
\begin{aligned}\label{eq: weyl_folding}
    &\frac{1}{|W|}\sum_{\lambda\in \Lambda_W/(k+N) \Lambda_R}\left(\sum_{w\in W}\theta_{w(\lambda + \rho)}(q, x)\det(w)\right)\left(\sum_{w'\in W}\theta_{w'(\lambda + \rho)}(\bar{q}, y)\det(w')\right)\nonumber\\
    =&\frac{1}{|W|}\sum_{\lambda\in \Lambda_W/(k+N) \Lambda_R}\left(\sum_{w\in W}\theta_{w(\lambda)}(q, x)\det(w)\right)\left(\sum_{w'\in W}\theta_{w'(\lambda)}(\bar{q}, y)\det(w')\right)\nonumber\\
    = &\sum_{\substack{\lambda \in \Lambda_W/(k + N)\Lambda_R\\w \in W}}\theta_{\lambda}(q, x)\theta_{w(\lambda)}(\bar{q}, y)\det(w)\nonumber\\
    = &\sum_{\substack{\lambda \in \Lambda_W/(k + N)\Lambda_R\\w \in W}}\theta_{\lambda}(q, x)\theta_{\lambda}(\bar{q}, w(y))\det(w)
\end{aligned}
\end{equation}
The first equality relies on the fact that $\lambda\to \lambda + \rho$ is an isomorphism of $\Lambda_W/(k + N)\Lambda_R$. The second equality follows from a change of variable $w' = \kappa w, w(\lambda) = \lambda'$. The sum over $w$ cancels $1/|W|$. The last equality follows from standard symmetries of the theta functions (appendix \ref{app: convention}).

In the next section, we will perform a Poisson resummation of the lattice sum $Z_{\text{susy}}$ and shows that it reproduces this sum of products of theta functions.

\subsection{A Partial Poisson Resummation}\label{sec:poisson}
In this section, we perform a Poisson resummation of $Z_{\text{susy}}$. First, we set $w = 1$ and complete the square in the exponential
\begin{align*}
    &\exp(-\frac{\pi(k + N)}{2\tau_2}((n - s\bar{\tau} - 2 \alpha_L , n - s\tau + 2 \alpha_R) -2i\tau_2(n, s)
    + (\alpha_L + \alpha_R)^2))\\
    =&
    \exp(-\frac{\pi(k + N)}{2\tau_2}\left((n - s\tau_1 - \alpha_L + \alpha_R - i\tau_2s)^2 + 2\tau_2^2s^2 - 2is^2\tau_1\tau_2 + 4i\alpha_R\tau_2s\right))
\end{align*}
We now use the following Poisson resummation formula\footnote{The formula works more generally if we replace $\Lambda_R$ by an arbitrary full-dimensional lattice in an inner product space. It is first proved for $x$ real and then we perform analytic continuation.} for $A > 0$ and $x$ an arbitary complex vector of size $N$:
\begin{eqnarray}
    \sum_{n\in \Lambda_R} \exp(- A(x + n)^2) = \sqrt{\frac{\pi}{A}}^{\dim \Lambda_R}\frac{1}{|\Lambda_R^*|}\sum_{m\in \Lambda_R^*}\exp(-\frac{\pi^2m^2}{A})\exp(2\pi i (m, x))
\end{eqnarray}
where $\Lambda_R$ is the root lattice of $SU(N)$ and its dual lattice $\Lambda_R^* = \Lambda_W$ is the weight lattice of $SU(N)$. We use the convention that the inner products of a lattice and its dual are integers. $\dim \Lambda_R = N - 1$ is the dimension of $\Lambda_R$ and $|\Lambda_R^*|$ is the volume of the unit cell of $\Lambda_R^*$. Now we perform a Poisson resummation with respect to $n$:
\begin{equation}
\begin{aligned}
    &Z_{\text{susy}}(\tau, \bar{\tau}, \alpha_L, \alpha_R) \\
     =& \sum_{\substack{w\in W\\n, s\in \Lambda_R}}\det(w)\sqrt{\frac{2\tau_2}{k + N}}^{1 - N}|\Lambda_R^*|\\
    &\quad\quad\times\exp(-\frac{\pi(k + N)}{2\tau_2}((n - s\bar{\tau} - 2 w(\alpha_L) , n - s\tau + 2 \alpha_R) -2i\tau_2(n, s)
    + (w(\alpha_L) + \alpha_R)^2))
    \\
    = & \sum_{\substack{w\in W\\m\in \Lambda_W\\s \in \Lambda_R}}\det(w)\exp(-\frac{2\pi \tau_2m^2}{k + N})\exp(2\pi i (m, -w(\alpha_L) + \alpha_R - s\tau))\\
    &\quad\quad\times\exp(-\pi(k + N)\tau_2s^2 + \pi(k + N)s^2\tau_1i - 2\pi(k + N)i(\alpha_R, s))\\
     = &\sum_{\substack{w\in W\\m\in \Lambda_W\\s \in \Lambda_R}}\det(w) x^{m - (k + N)s}w(y)^{m}q^{(m - (k + N)s)^2/(2(k  + N))}\bar{q}^{m^2/(2(k + N))}\\
    = &\sum_{\substack{w\in W\\ \lambda \in \Lambda_W/(k + N)\Lambda_R}} \det(w)\theta_\lambda (q, x)\theta_\lambda(\bar{q}, w(y))
\end{aligned}
\end{equation}
where we have used the definitions
\begin{eqnarray*}
    q = \exp(2\pi i \tau)\quad \bar{q} = \exp(-2\pi i \bar{\tau})\quad x_i = \exp(2\pi i \alpha_R^i) \quad y_i = \exp(-2\pi i \alpha_L^i)
\end{eqnarray*}
The last equality follows from the following stronger result:
\begin{align*}
    &\sum_{\substack{m\in \Lambda_W\\s \in \Lambda_R}} x^{m - (k + N)s}y^{m}q^{(m - (k + N)s)^2/(2(k  + N))}\bar{q}^{m^2/(2(k + N))}\\
    = &\sum_{\substack{ \lambda \in \Lambda_W/(k + N)\Lambda_R}} \theta_\lambda (q, x)\theta_\lambda(\bar{q}, y)
\end{align*}
The stronger version implies the weaker version by summing over the Weyl group. To prove the stronger version, we need to check that the set of ordered pairs:
\begin{eqnarray}
    (m - (k + N)/2, m)\quad m \in \Lambda_W\quad s\in \Lambda_R
\end{eqnarray}
equals the set of ordered pairs:
\begin{eqnarray}
    (\mu + (k + N)r, \mu + (k + N)r')\quad \mu\in \Lambda_W\quad r,r'\in \Lambda_R
\end{eqnarray}
This can be easily checked. The proof of the equality of $Z_{\text{susy}}$ and \eqref{eq: weyl_folding} is now complete. We showed in the previous section and \eqref{eq: weyl_folding} equals the left hand side of \eqref{eq: conjecture}. Hence the proof of the Murthy-Witten conjecture is complete.

\acknowledgments
We thank Paul Luis Rhoel, Panos Betzios, Noah Porcelli, Philip Boyle Smith, Ida Zadeh, Matthias Gaberdiel, Ron Reid-Edwards, Alex Colling, Bob Knighton, Zhipu Zhao for helpful discussions. We thank Panos Betzios for reading through the draft and providing helpful comment. This work has been partially supported by STFC consolidated grant ST/X000664/1.

\appendix
\section{Convention} \label{app: convention}
\begin{enumerate}
    \item We study $SU(N)$ WZW models at level $k$ throughout this paper. The letter $N$ and $k$ always have the same meaning and we suppress the dependence of various objects on them in the notations.
    \item $\Lambda_W$ is the weight lattice of $SU(N)$ and consists of vectors of the form:
    \begin{eqnarray}
        \mu = (\mu_1, \mu_2..., \mu_N)\quad \sum_i \mu_i = 0\quad \mu_i - \mu_j\in\Z, \forall i, j
    \end{eqnarray}
    A weight $\mu$ corresponds to the linear functional $(z_1,..., z_N)\mapsto \mu_1 z_1 + \mu_2 z_2+... + \mu_N z_N$ where $z_i$ are the diagonal entries of the Cartan subalgebra of $SU(N)$. 
    \item $\Lambda_R$ is the root lattice of $SU(N)$ and consists of vectors of the form:
    \begin{eqnarray}
        r = (r_1,r_2 ..., r_N) \quad \sum_i r_i = 0\quad r_i\in \Z, \forall i
    \end{eqnarray}
    \item $SU(N)$ fugacities are of the form:
    \begin{eqnarray}
        x = (x_1,..., x_N) \quad \prod_i x_i = 1
    \end{eqnarray}
    They are the diagonal entries of the maximal torus of $SU(N)$.
    In the main text we have two sets of fugacities $x_i$ and $y_i$, they are related to complex vectors $\alpha_L, \alpha_R$ via:
    \begin{eqnarray}
        x_i = \exp(2\pi i \alpha_R^i)\quad y_i = \exp(-2\pi i \alpha_L^i)
    \end{eqnarray}
    \item If $\mu = (\mu_1,..., \mu_N)\in \Lambda_W$ is a weight, the notation $x^\mu$ means
    \begin{eqnarray}
        x_1^{\mu_1}x_2^{\mu_2}... x_N^{\mu_N}
    \end{eqnarray}
    \item All inner products are the usual Euclidean inner product
    \begin{eqnarray}
        (\mu, \nu) = \mu_1\nu_1 + \mu_2\nu_2+... + \mu_N\nu_N
    \end{eqnarray}
    In particular, the weight lattice is dual to the root lattice in the sense that the inner product of any weight and any root is an integer. 
    \item $W\cong S_N$ is the Weyl group of $SU(N)$ and naturally permutes any vector of length $N$. If $\mu$ is a vector of length $N$ (e.g. a weight), we write $w(\mu)$ for the action of $w$ on $\mu$. $|W| = N!$ is the size of the Weyl group. If $w\in W$, $\det(w)$ is the determinant of $w$ as an orthogonal linear map on the Cartan subalgebra of $SU(N)$. It equals $\text{sgn}(w)$ in \cite{MurthyWitten}. 
    \item The summation
    \begin{eqnarray}
        \sum_{\nu\in \text{Integrable}}
    \end{eqnarray}
    is over the set of integrable $SU(N)$ weights at level $k$. An integrable weight $\nu \in \Lambda_W$ satisfies the additional constraint:
    \begin{eqnarray}
        \nu_1 - \nu_N\leq k
    \end{eqnarray}
    \item The Weyl vector for $SU(N)$ is
    \begin{eqnarray}
        \rho = \left(\frac{N - 1}{2}, \frac{N - 3}{2},..., -\frac{N-1}{2}\right)
    \end{eqnarray}
    The gap between any two adjacent elements of $\rho$ is one.
    \item The theta functions are defined as:
    \begin{eqnarray}
        \theta_{\mu}(q, x) = \sum_{r\in \Lambda_R}x^{\mu + (N + k)r}q^{(N + k)(r + \mu/(N + k))^2/2}
    \end{eqnarray}
    where $\mu\in \Lambda_W$ is a weight of $SU(N)$ and $r$ sums over the root lattice of $SU(N)$. For example, take $N = 3, k = 2$ with fugacities $x_1, x_2, x_3$ such that $x_1x_2x_3 = 1$.
    \begin{align*}
        &\mu = \left(-\frac{2}{3},\frac{1}{3}, \frac{1}{3}\right), r = (1, -1, 0)\Rightarrow \\
        &x^{\mu + (N + k)r}q^{(N + k)(r + \mu/(N + k))^2/2} = x_1^{13/3}x_2^{-14/3}x_3^{1/3} q^{((13/3)^2 + (14/3)^2 + (1/3)^2)/10}
    \end{align*}
    This theta function is invariant under $\mu\to \mu + (k + N) \Lambda_R$. Hence the domain of $\mu$ can be regarded as $\Lambda_W/(k + N)\Lambda_R$. The theta function also satisfies the following relation:
    \begin{eqnarray}
        \theta_{w(\mu)}(q, x) = \theta_{\mu}(q, w^{-1}(x))\quad \theta_{-\mu}(q, x) = \theta_{\mu}(q, x^{-1})
    \end{eqnarray}
    where $w\in W$ is an element of the Weyl group.
\end{enumerate}

\section{Shifted Weyl Action on $\Lambda_W/(k + N)\Lambda_R$}\label{sec:shifted_weyl}
In this section, we prove\footnote{We thank Zhipu Zhao for discussions on this proof.} the two claims \ref{enum: shifted_weyl_action} on the shifted Weyl action on $\Lambda_W/(k + N)\Lambda_R$. Let $\mu\in \Lambda_W$ be an $SU(N)$ weight. We can subtract $\mu$ by an element inside $(k + N)\Lambda_R$ until the maximal difference between any two components of $\mu$ (always nonnegative by definition) is at most $k + N$: if the maximal difference is achieved at $\mu_i - \mu_j > N + k$. We can subtract $\mu_i$ by $N + k$ and add to $\mu_j$ by $N + k$. The maximal difference decreases after this step. So this algorithm terminates. Now
there are three possibilities:
\begin{enumerate}
    \item The maximum difference between components of $\mu$ is exactly $k + N$. In this case, swapping the two components with the maximal difference is a Weyl transformation of determinant $-1$ that fixes $\mu$ up to $(k + N)\Lambda_R$.
    \item The maximum difference between the components of $\mu$ is strictly less than $k + N$ and the components are not distinct. Then we can swap the nondistinct elements.
    \item The maximum difference between the components is strictly less than $k + N$ and the elements are distinct. Then we use a Weyl transformation to map it to the fundamental chamber and subtract $\rho$ from it to obtain an integrable weight.
\end{enumerate}
To see that integrable weights have trivial stabilizers under the shifted Weyl action, assume that \begin{eqnarray}
	\lambda_i  + \rho_i - w(\lambda + \rho)_i = (k + N) r_i\quad r_i\in \Z\quad \sum_{i = 1}^N r_i = 0\quad w\neq 1
\end{eqnarray}
where $\lambda$ is integrable. Since $\lambda + \rho$ have pairwise distinct components, $r$ must be nonzero.
A nonzero element of the root lattice must have both positive and negative components: $r_i > 0, r_j < 0$ for some $i\neq j$. The maximum difference between components of $\lambda + \rho$ is at most $k + N - 1$. Hence, the difference between any two components of $\lambda  + \rho - w(\lambda + \rho)$ is at most $2(k + N - 1)$ which contradicts $(k + N)(r_i - r_j)\geq 2(k + N)$.
\bibliographystyle{JHEP}
\bibliography{WZW_Pf}

\end{document}